\documentclass[mathleft
]{an}
\usepackage{graphicx}
\usepackage{times}
\usepackage{color}
\usepackage[sort]{natbib}

\usepackage[hidelinks=true]{hyperref}
\definecolor{dark-red}{rgb}{0.4,0.15,0.15}
\definecolor{dark-blue}{rgb}{0.1,0.1,0.6}
\definecolor{medium-blue}{rgb}{0,0,0.5}
\hypersetup{
    colorlinks, linkcolor={dark-red},
    citecolor={dark-blue}, urlcolor={medium-blue}
}
\def\aap{A\&A}
\def\solphys{Sol.~Phys.}
\def\aapr{A\&ARv}
\def\nat{Nature}
\def\pasp{PASP}

\def\apj{ApJ}
\def\apjl{ApJ}

\begin{document}

\Pagespan{000}{}
\Yearpublication{}%
\Yearsubmission{}%
\Month{}%
\Volume{}%
\Issue{}%
\DOI{\today} 

\title{The association between sunspot magnetic fields and superpenumbral fibrils}

\author{R.E. Louis\inst{1}\fnmsep\thanks{Corresponding author:
  \email{rlouis@aip.de}\newline}
\and  H. Balthasar\inst{1}
\and  C. Kuckein\inst{1}
\and  P. G$\ddot{\textrm{o}}$m$\ddot{\textrm{o}}$ry\inst{2}
\and  K.G. Puschmann\inst{1}
\and  C. Denker\inst{1}
}
\titlerunning{The association between sunspot magnetic fields and superpenumbral fibrils}
\authorrunning{Louis et al.}
\institute{
Leibniz-Institut f$\ddot{\textrm{u}}$r Astrophysik Institut Potsdam (AIP), An der Sternwarte 16, 
D-14482 Potsdam, Germany
\and 
Astronomical Institute, Slovak Academy of Sciences, SK-05960 Tatransk\'a Lomnica, Slovakia
}
 
\received{}
\accepted{}
\publonline{}


\abstract{Spectropolarimetric observations of a sunspot were carried out with the Tenerife Infrared Polarimeter at 
Observatorio del Teide, Tenerife, Spain. Maps of the physical parameters were obtained from an inversion of the 
Stokes profiles observed in the infrared Fe {\sc i} line at 15648~\AA. The regular sunspot consisted 
of a light bridge which separated the two umbral cores of the same polarity. One of the arms of the light bridge 
formed an extension of a penumbral filament which comprised weak and highly inclined magnetic fields. In addition, 
the Stokes $V$ profiles in this filament had an opposite sign as the sunspot and some resembled Stokes $Q$ or $U$. 
This penumbral filament terminated abruptly into another at the edge of the sunspot, where the latter was relatively 
vertical by about 30$^\circ$. Chromospheric H$\alpha$ and He 304\AA~filtergrams revealed three superpenumbral fibrils 
on the limb-side of the sunspot, in which one fibril extended into the sunspot and was oriented along the highly 
inclined penumbral counterpart of the light bridge. An intense, elongated brightening was observed along this fibril 
that was co-spatial with the intersecting penumbral filaments in the photosphere. Our results suggest that the disruption 
in the sunspot magnetic field at the location of the light bridge could be the source of reconnection that led to the 
intense chromospheric brightening and facilitated the supply of cool material in maintaining the overlying superpenumbral 
fibrils.}

\maketitle

\section{Introduction}
\label{intro}
Sunspots are sites of strong magnetic fields. Sunspots comprise a central dark core 
called the umbra, which is surrounded by a relatively, brighter, radially-oriented, filamentary structure called the 
penumbra. Magnetic fields are strongest in the umbra, exceeding 2.5 kG and are almost 
vertical \citep[][and references therein]{2003A&ARv..11..153S,2008A&A...488.1085B,2011LRSP....8....4B}.
The magnetic field inclination to the 
vertical increases with increasing radius, reaching an average value of 70$^\circ$ at the edge of the spot, where the 
field strength drops to less than 1000 G \citep{2004A&A...427..319B}. The filamentation of the penumbra is also reflected 
in its magnetic field, where stronger and relatively vertical magnetic fields alternate with filaments that are weaker and 
more inclined \citep{2008A&A...481L..13B,2010ApJ...721L..58P, 2010ApJ...720.1417P}. The photospheric magnetic field extends 
beyond the visible sunspot boundary in the form of a low-lying superpenumbral canopy \citep{1994A&A...283..221S}. 

In the chromosphere, the outermost boundaries of sunspots are referred to as superpenumbrae which comprise thin, dark, 
slightly curved fibrils or filaments, that extend radially outwards from the sunspot. According to 
\citet{1925PASP...37..268H,1927Natur.119..708H} and \citet{1941ApJ....93...24R}, the vorticity of superpenumbral fibrils 
follows a hemispherical rule wherein most sunspots in the northern/southern hemisphere comprised counterclockwise/clockwise 
fibrils. \citet{2004ApJ...608.1148B} reported that clockwise and counterclockwise fibrils could typically exist in the 
same superpenumbra with one-third of them originating from inside the penumbra. They concluded that the topology of fibrils 
was affected by the distribution of magnetic fields around the sunspot \citep{1996MNRAS.278..821P}. 

In this paper, we analyze the association of superpenumbral fibrils in the chromosphere with the photospheric magnetic field 
of a regular, unipolar sunspot. We find evidence for strong variations in the magnetic field in the vicinity of a light bridge 
that are possibly related to the dynamics observed in the chromosphere.

\section{Observations}
\label{obs}
\begin{figure}[!h]
\includegraphics[angle=90,width = \columnwidth]{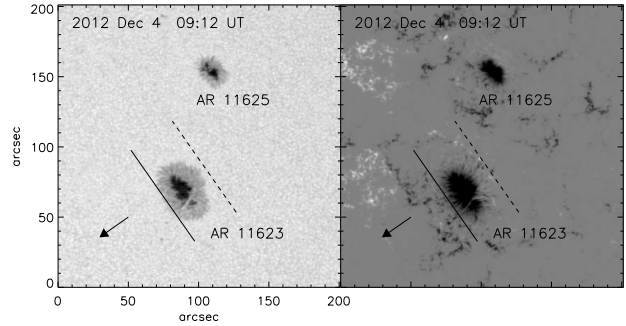}
\vspace{-20pt}
\caption{Slit orientation and scan direction of TIP II. The left and right panels correspond to the HMI continuum intensity 
filtergram and line-of-sight magnetogram, respectively. The initial and final slit positions are indicated by the solid and 
dashed lines, respectively. The arrow points to disc center. Solar east and north are to left and top, respectively.}
\label{fig01}
\end{figure}

We carried out spectropolarimetric observations of active region NOAA 11623 on 2012 December 4 using the Tenerife Infrared 
Polarimeter \citep[TIP II;][]{2007ASPC..368..611C} at the German Vacuum Tower Telescope (VTT), Observatorio del Teide, Tenerife, 
Spain. TIP II provided spectral scans in the two infrared Fe~{\sc i} lines (15648~\AA, $g_{eff} = 3$ and 15652~\AA, $g_{eff} = 1.53$) 
with full-Stokes polarimetry. The sunspot was located at N7/W12 and at a heliocentric angle of 14$^\circ$. At each slit position 
and for a single Stokes parameter, we accumulated ten exposures of 250~ms each. Using a 
0\farcs36-wide slit, a full scan of the sunspot was carried out with 100 scan steps, each step being 0\farcs36. The scan lasted 
from 9:08--9:29 UT. The spatial sampling along the slit was 0\farcs175 while the spectral sampling was 19.8~m\AA. The scan was 
facilitated by an incrementally turned dichroic beam splitter plate placed in the adaptive optics tank on the first floor of 
the VTT. This beam splitter transmits a small fraction of light to the wavefront sensor (WFS) and reflects all other light to the 
science focus. A small tilt shifts the image at the science focus but not on the WFS. Figure~\ref{fig01} shows the orientation of 
the spectrograph slit of TIP II, overlaid on a continuum intensity filtergram and line-of-sight magnetogram from the Helioseismic 
and Magnetic Imager \citep[HMI;][]{2012SoPh..275..229S}. The initial and final slit positions are indicated by the solid and dashed 
lines, respectively. Flat-fielding and polarization calibration were performed at the end of the scan. This was done using a rotatable 
linear polarizer and phase plate which are introduced in the optical path after the exit window of the vacuum tower \citep{1999ASPC..184....3C}.

\begin{figure}[!h]
\centerline{
\includegraphics[angle=0,width = 1.0\columnwidth]{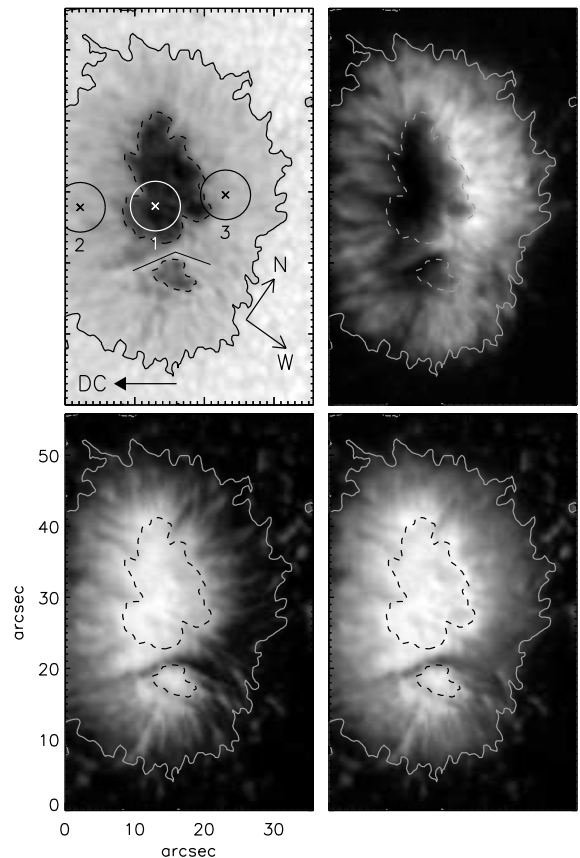}
}
\vspace{-5pt}
\caption{TIP II scan of the leading sunspot in active region NOAA 11623. Clockwise from top left: Continuum intensity at 15648~\AA, 
total linear polarization, total polarization, and total circular polarization. The {\em cross} symbols indicate pixels whose Stokes 
profiles are shown in Fig.~\ref{fig03}. Pixels numbered 1, 2, and 3 correspond to the umbra, center--side, and limb--side penumbra, 
respectively. DC--disc center, N--solar north, and W--solar west.}
\label{fig02}
\end{figure}

Our observations were supported by the Kiepenheuer-Institut Adaptive Optics System \citep[KAOS;][]{2003SPIE.4853..187V} which enabled 
us to obtain a highly stable scan of the sunspot with a spatial resolution of about 0\farcs8. Figure~\ref{fig02} shows a slit-reconstructed 
image of the sunspot in continuum intensity (top left panel) at 15648~\AA~from the TIP II scan. The sunspot comprises of two umbrae separated 
by a large light bridge (LB). The LB is oriented in the shape of an inverted `\textsf{V}' (marked in the upper left panel of Fig.\ref{fig02} by
a thin black line) in which the smaller umbral core is nestled in the 
bow of the LB. The filamentation in the penumbra is clearly discernible in the maps of total linear (top right) and total circular 
polarization (bottom left). In comparison to the umbra and the center-side penumbra, the total circular polarization is weaker in 
the mid and outer regions of the limb-side penumbra as well as in the LB. The total polarization (bottom right) in the LB is 
relatively weaker as compared to other regions of the sunspot.

We also use full-disk H$\alpha$ images from the Global Oscillations Network Group \citep[GONG;][]{1996Sci...272.1284H,2011SPD....42.1745H} 
recorded at the Udaipur station in India. The 2k$\times$2k images have a spatial sampling of about 1\arcsec and a cadence 
of 1 min. These are supported by He {\sc ii} 304 \AA~images from the Atmospheric Imaging Assembly \citep[AIA;][]{2012SoPh..275...17L} 
which cover the chromosphere and transition region. The 4k$\times$4k-pixel images have a spatial sampling and cadence of 
0\farcs6 and 2 min, respectively.

\begin{figure}[!h]
\includegraphics[angle=0,width = \columnwidth]{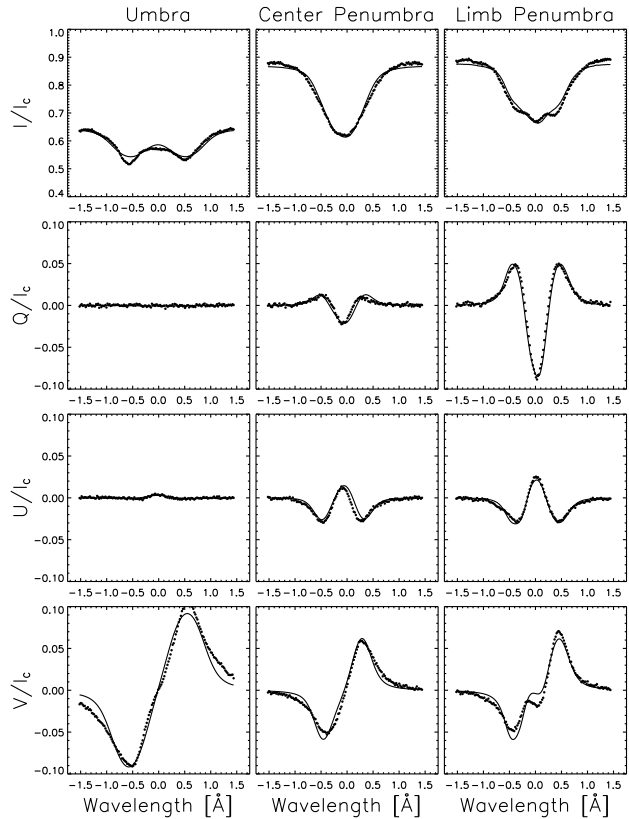}
\vspace{-20pt}
\caption{Observed ({\em filled circles}) and synthetic ({\em solid line}) Stokes profiles from SIR for the three 
pixels indicated in Fig.~\ref{fig02}. From left to right--umbra, center--side, and limb--side penumbra.}
\label{fig03}
\end{figure}
\section{Results}
\subsection{Photospheric magnetic field}
\label{mag}
The infrared Fe lines are very sensitive to the magnetic field owing to the $\lambda^2$-dependence on the Zeeman splitting and 
their high Land\'e factors. In addition, these lines sample the deep photospheric layers due to the reduced opacity of H$^{-}$ 
at these wavelengths \citep{2004A&A...427..319B}. There are several ways of selecting the infrared Fe lines for inversion, 
which depends on the spatial position in the sunspot and/or the availability of additional spectral lines. In the umbra, 
the Fe {\sc i} 15652 \AA~line is blended with two molecular OH lines which show up in the Stokes $I$ and $V$ but are absent 
in $Q$ and $U$. \citet{2003A&A...410..695M} derived the three dimensional structure of a sunspot by inverting both the 
Fe lines at 15648 \AA~and 15652 \AA~as well as the molecular OH lines. \citet{2004A&A...427..319B} combined a third 
Fe {\sc i} line at 15647 \AA~with the above lines but did not consider the Stokes $I$ and $V$ profiles of the 15652 \AA~line 
in their inversions of sunspot umbra. \citet{2007A&A...475.1067C,2008A&A...477..273C} and 
\citet{2008A&A...480..825B,2011A&A...525A.133B} have demonstrated that combining the infrared lines along with the visible 
Fe lines at 6300~\AA, provides a better physical stratification as the different lines sample a large range of atmospheric layers. 
Nevertheless, it depends on the applied method how fast the magnetic field decreases with height 
\citep[see][]{2008A&A...488.1085B,2013CEAB...37..435B}.

The observed Stokes profiles from the TIP II scan were subjected to an inversion using the SIR \citep[Stokes Inversion based 
on Response functions;][]{1992ApJ...398..375R} code to retrieve the thermal, magnetic and kinematic parameters in the sunspot. A 
single magnetic component was assumed in each pixel. Temperature was perturbed with five nodes while two nodes were provided 
for the magnetic field and line-of-sight (LOS) velocity. Inclination and azimuth were assumed to be height-independent. The 
average quiet Sun (QS) profile was used to retrieve the fraction of stray--light which was also a free parameter in the 
inversion. After several attempts, we reverted to the inversion of the stronger Fe {\sc i} line at 15648 \AA~as the resulting 
fit to the observed profiles were inferior when performing the inversion with both lines simultaneously. As both of these lines 
have similar heights of formation, the resulting model atmosphere from a single line inversion should adequately 
describe the physical conditions at that height range.

\begin{figure*}[!ht]
\includegraphics[angle=0,width = \textwidth]{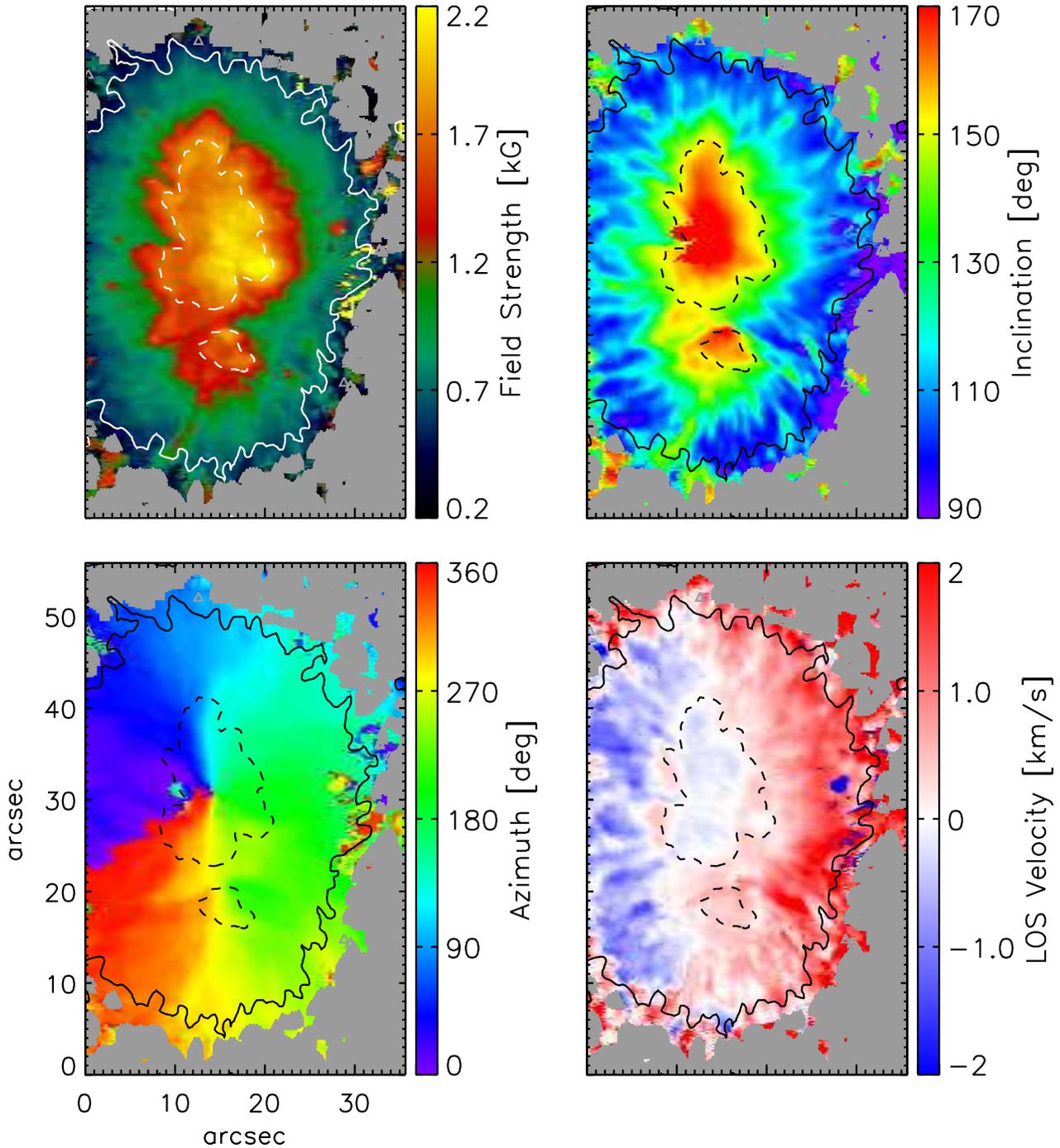}
\vspace{-25pt}
\caption{Maps of physical parameters retrieved from SIR. Clockwise from top left: field strength, inclination, 
LOS velocity, and azimuth. The field strength and LOS velocity correspond to the optical depth at $\log{\tau}=-0.5$. All 
images have been scaled to their respective colour bar. The field inclination and azimuth are expressed in the local reference 
frame, but are not deprojected. An inclination of 0$^\circ$/180$^\circ$ refers to the magnetic field pointing vertically up/below 
the solar surface. 0$^\circ$ azimuth is directed along the positive $x$--axis and increases in the clockwise direction. Blue 
and red colours in the LOS velocity map correspond to blue-shifts and red-shifts, respectively.}
\label{fig04}
\end{figure*}

Figure~\ref{fig03} shows the synthetic profiles from the SIR inversion for an umbral, center-side and limb-side penumbral 
pixel, respectively. Note the large Zeeman splitting seen in Stokes $I$ for the umbral pixel. The agreement between the 
observed and synthetic profiles thus validates our choice of the single component model atmosphere. Maps of the magnetic 
field strength, inclination, azimuth, and LOS velocity retrieved from the inversion are shown in Fig.~\ref{fig04}. The field 
strength and LOS velocity correspond to the optical depth at $\log{\tau}=0$. The inclination and azimuth have been transformed 
to the local reference frame (but not deprojected). The 180$^\circ$ ambiguity in the azimuth was removed manually with 0$^\circ$ 
azimuth directed along the positive $x$--axis and increases in the clockwise direction. Apart from a single bad patch 
($x=28^{\prime\prime}$, $y=31.5^{\prime\prime}$) in the outer limb-side penumbra, the maps of the physical parameters are very 
smooth and reveal small--scale structuring seen earlier in Fig.~\ref{fig02}. In the umbra, the maximum and mean field strengths 
are 2200 and 1850~G at an optical depth of $\log{\tau}=-0.5$, respectively. The response of Stokes $I$ and $V$ is most 
sensitive to the magnetic field strength and LOS velocity at this height \citep{2003A&A...410..695M,2005A&A...439..687C} for this spectral line.
The field reduces from 1750 G at the umbra-penumbra 
boundary to about 600 G in the outer penumbra. In the LB, field strengths vary from 900--1400~G. 
While the magnetic field is predominantly vertical in the umbra, in the penumbra we find a variation of about 15$^\circ$ between 
adjacent penumbral filaments. The magnetic field is largely vertical (150$^\circ$) on the linear part of the LB (left half), while it 
is highly inclined (100$^\circ$) on the other anchorage section (right half) which is seen as an extension of the penumbra. The 
Evershed flow \citep[EF;][]{1909MNRAS..69..454E} at $\log{\tau}=0$ is about $\pm$1.5~km~s$^{-1}$ on the center-side and limb-side penumbra, 
respectively. However, at some locations in the outer limb-side penumbra, the LOS velocity exceeds 3~km~s$^{-1}$.

\begin{figure*}[!ht]
\centerline{
\includegraphics[angle=90,width = \textwidth]{halpha-image.eps}
}
\vspace{-5pt}
\caption{Chromospheric H$\alpha$ image of active region. The inset in panel `a' is a magnified view of the sunspot in 
active region 11623, scaled to the TIP II scan. The white contours represent the photospheric continuum intensity at 
15648~\AA. Panels `b' and `c' show a magnified view of the field inclination and H$\alpha$ intensity, respectively, for 
the sub-region marked by the black rectangle in the inset of panel `a'. The grey arrows represent the horizontal magnetic 
field. The horizontal grey arrow at the bottom of panel `c' corresponds to 1~kG.}
\label{fig05}
\centerline{
\includegraphics[angle=90,width = 0.78\textwidth]{overlay-stokes-v.eps}
}
\vspace{-10pt}
\caption{Intersection of a strongly horizontal and relatively vertical penumbral filament and their associated Stokes 
$V$ profiles. The profiles have been overlaid on the inclination which has been magnified by a factor of 3. The profiles are 
indicated for every pixel in the horizontal direction and every second pixel in the vertical direction corresponding to the original
image. The $y$--axis of the profiles has been clipped to $\pm$3\%.}
\label{fig06}
\end{figure*}

\subsection{Connection between sunspot light bridge and superpenumbral fibrils}
\label{fibril}
Panel `a' of Fig.~\ref{fig05} depicts the chromosphere of the active region in H$\alpha$. The sunspot in question comprises 
three distinct superpenumbral fibrils on its western side which are marked with arrows. There is also a closed bundle of 
fibrils south of the sunspot. Out of these three fibrils, the lower most (labeled `3') is seen to extend all the way to the LB although it 
is disjoint at the periphery of the sunspot. The inset in panel `a' is a magnified view of the sunspot in H$\alpha$ scaled 
to the TIP II continuum intensity map. It depicts a short part of the fibril that terminates at the edge of the sunspot. The 
foot-point separation and the length of the superpenumbral fibrils vary from 18--27~Mm. The fibrils do not have a distinct 
clockwise or counter-clockwise orientation and are seen as radially-orientated, elevated structures.

Panels `b' and `c' of Fig.~\ref{fig05} show a close-up view of the field inclination and H$\alpha$ intensity, respectively, 
for the sub-region marked by the black rectangle in the inset of panel `a'. The grey arrows represent the horizontal magnetic 
field. These panels reveal that a part of the chromospheric fibril coincides with the highly inclined penumbral filament that 
is a continuation of the LB. In the outer penumbra, this horizontal filament abruptly terminates with another filament which is relatively 
vertical. We estimate the difference in inclination at the location of intersection to be about 30$^\circ$. As the sunspot 
has negative polarity, the horizontal magnetic field is directed radially inwards and is seen to be aligned with the 
chromospheric fibril. Figure~\ref{fig06} shows the Stokes $V$ profiles associated with this pair of intersecting penumbral 
filaments with the inclination as the background image which has been magnified by a factor of 3. The more horizontal filament that 
coincides with the fibril in H$\alpha$, comprises profiles which have the opposite sign as that of the sunspot or resemble 
Stokes $Q$ or $U$. In addition, the signals are quite weak in comparison to those in the relatively vertical penumbral filament 
where the profiles consist of two anti-symmetric lobes, and have the same polarity as the sunspot.

\begin{figure}[!h]
\hspace{-10pt}
\centerline{
\includegraphics[angle=0,width = 1.07\columnwidth]{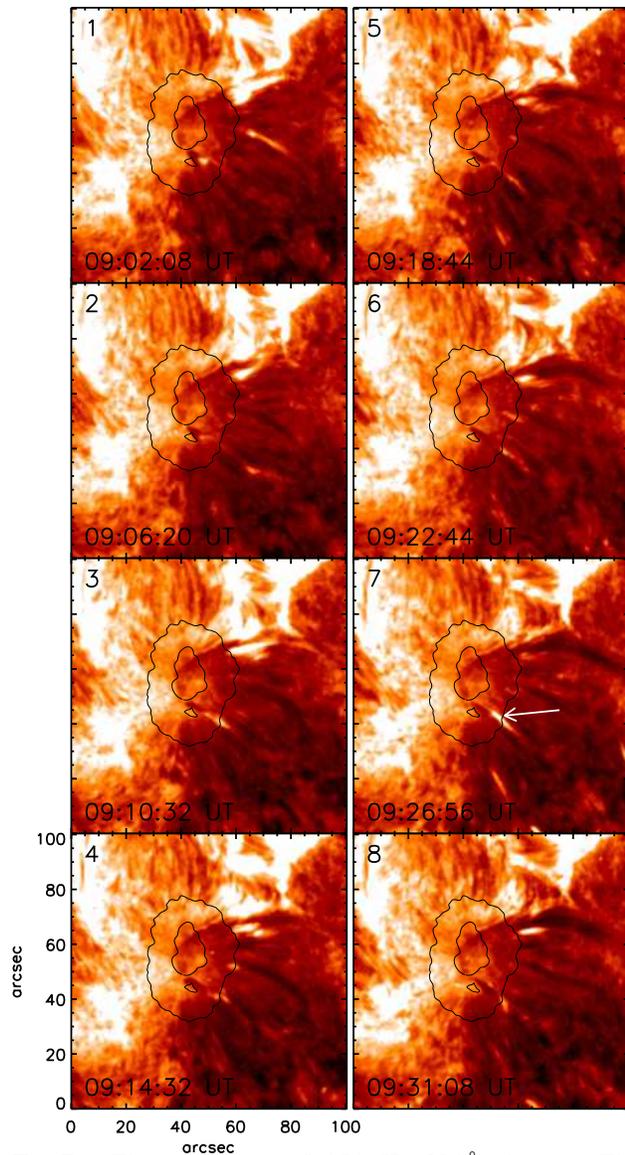}
}
\vspace{-30pt}
\caption{Time sequence of AIA He 304\AA~ images. The white arrow in panel 7 marks an intense brightening along one of 
the superpenumbral fibrils that extends to the sunspot light bridge.}
\label{fig07}
\end{figure}

Figure~\ref{fig07} shows the time sequence of AIA He {\sc ii} 304 \AA~images which indicate that the superpenumbral 
fibrils seen in H$\alpha$ originate inside the penumbra. The fibrils are also associated with strong emission in 
the periphery of the sunspot (panels 1 and 8). However, in the case of the lowermost fibril described above, we 
find an intense brightening that is located well within the sunspot and co-spatial with the intersecting penumbral 
filaments (white arrow in panel 7). The elongated brightening is a part of the fibril that originates at the 
sunspot LB, similar to the H$\alpha$ images. A movie of the H$\alpha$ and He {\sc ii} 304\AA~ sequence gives the 
impression of material moving radially inwards along the fibril. However, this could be attributed to the 
following scenarios: i) the motion is a true motion of plasma along the field lines which could be associated with the chromospheric 
inverse-EF \citep{1913ApJ....37..322S,1969SoPh....9...88H,1990A&A...233..207D,2003ApJ...584..509G}, and ii) the motion is apparently inward
which occurs when the plasma at different segments of the fibril reaches chromospheric temperatures.

\section{Summary and conclusions}
\label{conclu}
We carried out spectropolarimetric observations with TIP II at the VTT, Observatorio del Teide, Tenerife, Spain of a sunspot 
in active region NOAA 11623 on 2012 December 4. Spectral scans of the Fe {\sc i} line at 15648~\AA~were obtained with full 
Stokes polarimetry. The observed Stokes profiles were subjected to an inversion using the SIR code which yielded maps of the 
physical parameters. The sunspot was a unipolar, regular structure with a large LB which separated the two umbral cores of the 
same polarity. The LB was oriented in the shape of an inverted `\textsf{V}' with the smaller umbral core located in the bow of the LB. 
While one of the halves of the LB on the center-side of the sunspot revealed the presence of relatively strong and vertical 
magnetic fields, the other half on the limb-side was seen as an extension of the penumbra with relatively weaker and highly 
inclined magnetic fields. This highly inclined penumbral filament terminated abruptly into another filament that was relatively 
vertical by about 30$^\circ$. These intersecting filaments were located near the penumbra-QS boundary. An inspection of the 
Stokes $V$ profiles in the highly inclined filament indicated the presence of $Q$ or $U$-like profiles as well as with an
opposite sign as the sunspot. By comparison, the other filament comprised stronger, normal, and anti-symmetric profiles 
with the same sign as the sunspot. 

In H$\alpha$ and He 304 \AA~filtergrams, we identified three superpenumbral fibrils on the limb-side of the sunspot, in 
which one fibril extended into the sunspot and was oriented along the highly inclined penumbral counterpart of the LB. The 
horizontal magnetic field was seen to be directed along the fibril channel. Towards the end of the TIP II scan, the He 304\AA~images 
showed an intense, elongated brightening along this fibril that was also co-spatial with the intersecting penumbral filaments 
in the photosphere. The time sequence of H$\alpha$ and He {\sc ii} images revealed an apparent radial inward motion of plasma along the 
chromospheric fibrils although this could be related either to the chromospheric inverse-EF or 
to a transient heating of the material to chromospheric temperatures.

The LB is a region where the magnetic field is disturbed as a result of hot, convective plasma penetrating the sunspot 
from below \citep{2006A&A...453.1079J}. This perturbation could result in a complex arrangement of magnetic fields which 
is seen in the form of a highly inclined penumbral filament intersecting another where the magnetic field is more 
vertical. Similar magnetic configurations have been reported earlier where the LB consisted of irregular Stokes 
profiles \citep{2009ApJ...704L..29L,2013A&A...549L...4R}. A magnetic topology such as the above, could facilitate 
reconnection in the chromosphere where we find a bright patch in the He {\sc ii} 304\AA~images and is consistent 
with findings of \citet{2008SoPh..252...43L,2009ApJ...696L..66S,2009ApJ...704L..29L} and \citet{2011ApJ...727...49L}. It 
has been shown that magnetic reconnection in the chromosphere plays an important role in the formation of 
filaments \citep{1989ApJ...343..971V,2000SoPh..195..333C,2007ApJ...659.1713V,2009ApJ...697..913O,2013ApJ...763...97W}, as 
it re-orients magnetic field lines and facilitates the supply of cool material to the filament. We find that such a 
process could occur as a result of a disruption of the sunspot magnetic field in the proximity of the light bridge thereby 
producing the intense chromospheric brightening along the superpenumbral fibril.

\acknowledgements
The Vacuum Tower Telescope is operated by the Kiepenheuer-Institute for Solar Physics in Freiburg, Germany, at the 
Spanish Observatorio del Teide, Tenerife, Canary Islands. This work utilizes data obtained by the Global Oscillation 
Network Group (GONG) Program, managed by the National Solar Observatory, which is operated by AURA, Inc. under a 
cooperative agreement with the National Science Foundation. The data were acquired by instruments operated by the 
Big Bear Solar Observatory, High Altitude Observatory, Learmonth Solar Observatory, Udaipur Solar Observatory, Instituto 
de Astrof\'isica de Canarias, and Cerro Tololo Interamerican Observatory. HMI data are courtesy of NASA/SDO and the 
HMI science team. They are provided by the Joint Science Operations Center -- Science Data Processing at Stanford 
University. We thank Dr. Ravindra for carefully reading the manuscript and providing his suggestions. 
R.E.L. and C.D. were supported by grant DE 787/3-1 of the German Science Foundation (DFG). P.G. acknowledges the 
support from grant VEGA 2/0108/12 of the Science Grant Agency. We thank the anonymous referee for the useful comments.


\end{document}